\documentclass[runningheads]{llncs}
\usepackage[T1]{fontenc}
\usepackage{graphicx}

\usepackage{amsmath}
\usepackage{amsfonts}
\usepackage{amssymb}
\usepackage{upgreek} % For upright Greek letters

\usepackage{hyperref}
\usepackage[capitalise,nameinlink]{cleveref} % load after hyperref package
\crefname{figure}{Fig.}{Figs.} 
\Crefname{figure}{Figure}{Figures}
\crefname{table}{Table}{Tables}
\Crefname{table}{Table}{Tables}
\usepackage{subcaption}
\begin{document}
\title{Implicit Neural Representations for Registration of Left Ventricle Myocardium During a Cardiac Cycle}
\titlerunning{INRs for Registration of the LVmyo during a cardiac cycle}
% \author{Anonymous}
% \authorrunning{Anonymous}

\author{Mathias Micheelsen Lowes \inst{1}%\orcidID{0009-0003-0002-1645} 
\and
 Jonas Jalili Pedersen\inst{2}%\orcidID{0000-0002-9496-7655} 
 \and
 Bjørn S. Hansen\inst{1}%\orcidID{0009-0003-6670-326X} 
 \and
 Klaus Fuglsang Kofoed\inst{2}%\orcidID{0000-0001-9742-1554} 
 \and
 Maxime Sermesant\inst{3} 
 \and
 Rasmus R. Paulsen\inst{1}%\orcidID{0000-0003-0647-3215}
 }
\authorrunning{M. M. Lowes et al.}
\institute{Department of Applied Mathematics and Computer Science, Technical University of Denmark, Kgs. Lyngby, Denmark \\ \email{mmilo@dtu.dk}\and
The Heart Center, Rigshospitalet, University of Copenhagen, Copenhagen, Denmark \and
Inria Université Côte d’Azur, Nice, France
}
\maketitle              % typeset the header of the contribution
\begin{abstract}

Understanding the movement of the left ventricle myocardium (LVmyo) during the cardiac cycle is essential for assessing cardiac function. One way to model this movement is through a series of deformable image registrations (DIRs) of the LVmyo. 
Traditional deep learning methods for DIRs, such as those based on convolutional neural networks, often require substantial memory and computational resources. In contrast, implicit neural representations (INRs) offer an efficient approach by operating on any number of continuous points. 
This study extends the use of INRs for DIR to cardiac computed tomography (CT), focusing on LVmyo registration.
% We apply two approaches for the registration of the LVmyo through a cardiac cycle: sequential and non-sequential. 
% The sequential approach outperforms the non-sequential approach when evaluated on metrics based on the LVmyo, but not for propagating landmarks through the cardiac cycle (only annotations from one patient available).
To enhance the precision of the registration around the LVmyo, we incorporate the signed distance field of the LVmyo with the Hounsfield Unit values from the CT frames. This guides the registration of the LVmyo, while keeping the tissue information from the CT frames.
Our framework demonstrates high registration accuracy and provides a robust method for temporal registration that facilitates further analysis of LVmyo motion.

\end{abstract}
\section{Introduction}
The movement of the left ventricle myocardium (LVmyo) during the cardiac cycle is complex and difficult to accurately model. Understanding and modelling the movement  of the LVmyo is important for assessing cardiac function and diagnosing cardiac diseases.
Modelling the movement of the LVmyo can be viewed as a deformable image registration (DIR) task requiring several registrations over a single cardiac cycle. 

\begin{figure}[t]
    \centering
    \includegraphics[width=\textwidth]{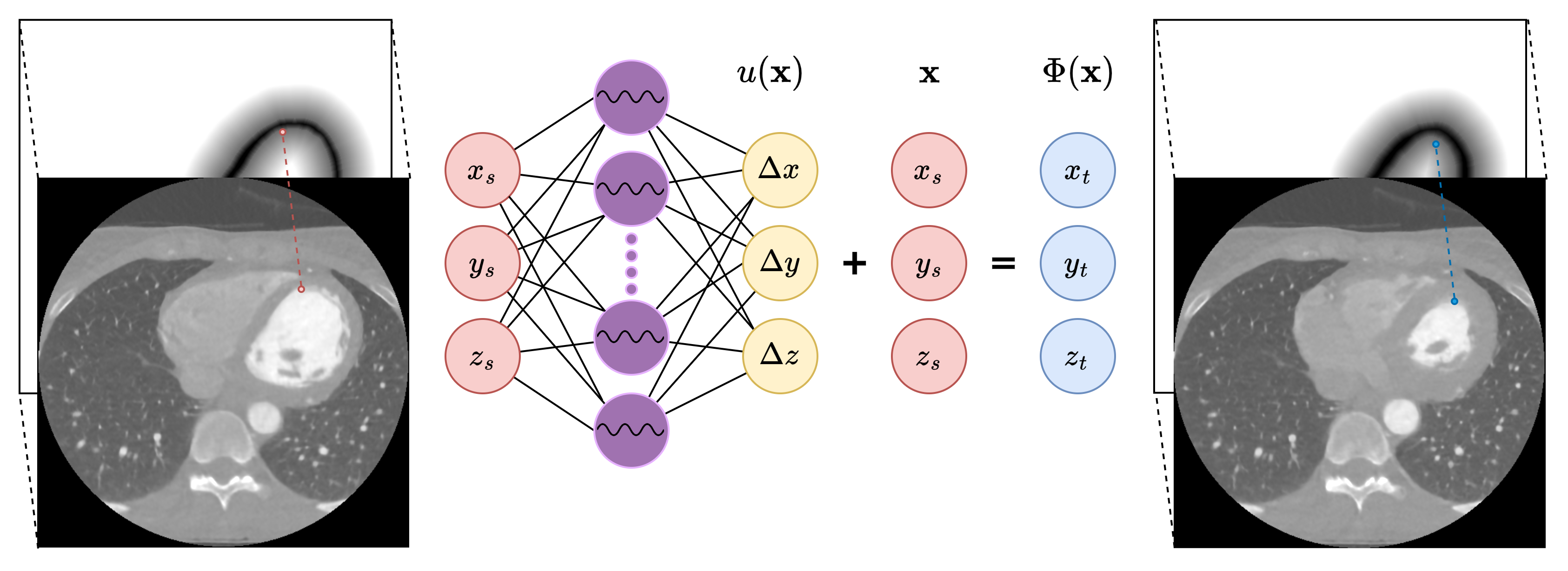}
    \caption{Registration process for a single point. The MLP learns the deformation $u(\mathbf{x})$ by sampling values from the CT frame and the SDF in the source domain at $\mathbf{x}$ and in the target domain at $\mathrm{\Phi}(\mathbf{x})$.}
    \label{fig:data_network}
\end{figure}

DIR involves establishing a correspondence between a source and a target image by deforming the source to align with the target. Often, the source image is mapped to a deformable vector field such that the location of each point is given by $\mathrm{\Phi}(\mathbf{x})=u(\mathbf{x})+\mathbf{x}$, where $\mathbf{x}$ is the position in the source image, see \cref{fig:data_network}. Applying $\mathrm{\Phi}$ to the coordinates of the source image aligns it to the target image.

Early approaches for DIR were mainly based on B-splines~\cite{Klein2010elastix,Rueckert1999FFD,Wu2016Bspline}. Through optimization, the parameters of the B-splines are calculated for the specific registration. Instead of being imposed to parameter optimization for each registration, deep learning approaches can be trained on a dataset and then generalize to new data.
Many deep learning approaches for DIR are based on convolutional neural networks (CNNs)~\cite{balakrishnan2019voxel,Heinrich2022voxelPlusPlus,Sokooti2017cnn,deVos2019cnnreg,Yang2017Patchbased}. 
CNNs offer the advantage of rapid inference post-training. However, they operate within the discrete voxel space of images, resulting in substantial memory demands. 
Consequently, downsampling the images is often necessary, which often compromise the registration accuracy.
% do down- and up-samlping similar to a gaussian image pyramid used with B-spline methods.
% While effective, these approaches are limited by the use of a discrete grid which requires a lot of memory, thus downsampling is often needed.

Implicit neural representations (INRs) address several limitations of using CNNs for DIR, primarily by operating on continuous points. This eliminates the restriction to a specific grid and reduces the memory requirements significantly. 
An INR leverages a neural network to approximate any signal or function. In the context of registration, the INR represents the function $\mathrm{\Phi}$, mapping a continuous input point to a continuous output point.
Wolterink et al.~\cite{wolterink2021implicit} demonstrated the ability of an INR to parameterize  a DIR between two 3D chest CT images, outperforming other learning based methods on the DIR-LAB dataset~\cite{Castillo2009DIRLAB} without the need for training data.
Several later works also explore and build on the use of INRs for registration \cite{Byra2023INRnature,harten2023deformable,vanharten2024cycle,Sun2022MIRNFMI,Tian2023NePhi}. Some of these works incorporate CNNs into the INR framework, achieving marginal performance improvement  while losing the simplicity of INRs and increasing the computational requirements.

Building on the framework of Wolterink et al.~\cite{wolterink2021implicit}, we extend the use of INR for DIR to the domain of cardiac CT with a specific focus on the registration of the LVmyo. Our work contributes by adding the signed distance field (SDF) of the LVmyo to the registration process. The SDF provide geometric information about the shape and position of the myocardium, which cannot be directly extrapolated from CT scans only using an INR.
By integrating both the HU values and the SDF values in the registration, our approach aims to capture the detailed and continuous movement of the LVmyo, improving the precision of the registration at each step.
%, and thereby improving understanding of myocardial motion.

% overall quality of the cardiac functional analysis.

% A deformable image registration is defined by the function $\mathrm{\Phi}(\mathbf{x})=u(\mathbf{x})+\mathbf{x}$, which takes a point $\mathbf{x}$ in the source domain and maps to the target domain. 
% By utilizing a multi-layer perceptron (MLP) as the neural network of the INR, the deformable registration ends up quite lightweight in comparison to CNN based approaches. Furthermore, the registration is not fixed to any grid points as it takes continuous coordinates as input.

% We aim to use INR's to represent the movement of the left ventricle myocardium during a heart cycle. We build on the frame work of Wolterink et al.~\cite{wolterink2021implicit} by including the signed distance field of the LVmyo in the registration phase. 
\section{Data}
% The data is acquired at Rigshospitalet, Copenhagen, Denmark as part of the ongoing Copenhagen General Population Study~\textbf{cite}. 
The data were acquired at Rigshospitalet, Copenhagen, Denmark, as part of the ongoing Copenhagen General Population Study.
It consists of 100 randomly selected participants from the study, who meet the requirements: individuals $>40$ years, non-pregnant and normal kidney function.
The participants underwent retrospective ECG-gated cardiac computed tomography angiography, allowing for continuous scanning of the heart over a cardiac cycle, resulting in one temporal scan per participant. 
Each temporal scan consists of 20 volumetric frames denoted $f_0,\ldots, f_{19}$, these are also divided into a percentage representation of the whole interval. The frame at $0\%$ ($f_0$) represents the end-diastole (ED) phase (maximum filling of the left ventricle), while the end-systole (ES) phase (maximum emptying of the left ventricle) usually occurs around $35-40\%$ of the series. 
The reconstructed spatial resolution of each frame is $0.5 \times 0.5 \times 2$ mm in the x-, y-, and z-directions, respectively.

In \cref{fig:data_network}, axial slices of two frames belonging to the same participant are displayed. The slice on the left is from the ED phase, while the slice on the right is from the ES phase.

\begin{figure}[htb]
    \centering
    \includegraphics[width=0.4\linewidth]{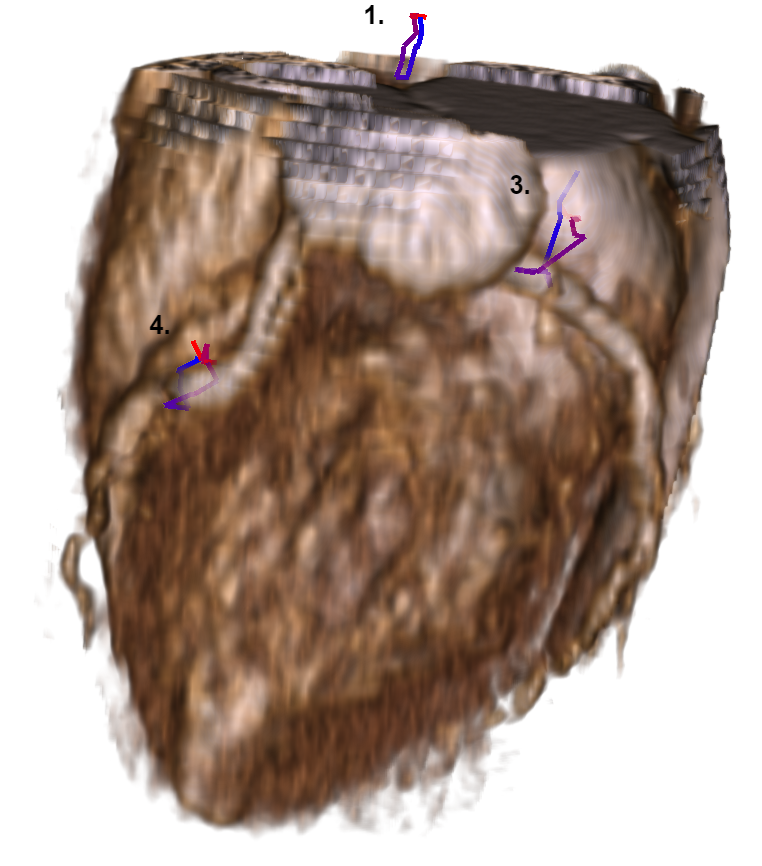}
    \caption{3D rendering of the CT frame in the ES-phase. The landmarks 1, 3 and 4 are rendered as lines representing the movement over a cardiac cycle, they are colored blue at 0\% with a gradient to red at 95\%.}
    \label{fig:lm_numbered}
\end{figure}

For a single participant, four expert-annotated landmarks are annotated on each CT frame in the cardiac cycle. These landmarks are defined from the coronary arteries as follows:

\begin{enumerate}
    \item The start of the left anterior descending artery (LAD).
    \item The start of the circumflex artery (CX).
    \item The point where the first obtuse marginal (OM) branches off the right circumflex artery.
    \item  The point where the last diagonal branches off the LAD.
\end{enumerate}
Landmarks 1, 3 and 4 are illustrated in \cref{fig:lm_numbered}.
Due to the high axial spacing of 2 mm, noise may be present in the landmarks since the exact point of the coronary artery could fall between two slices.

\section{Methods}

% The LVmyo registration in a cardiac cycle can be seen as a series of pairwise DIRs, each between a source image $I_S : \Omega \subset[-1,1]^d \rightarrow \mathbb{R}$ and a target image $I_T : \Omega \subset[-1,1]^d \rightarrow \mathbb{R}$. Each DIR aims to find a mapping $\mathrm{\Phi}: \mathbb{R}^d \rightarrow \mathbb{R}^d$ in the form of $\mathrm{\Phi}(\mathbf{x})=u(\mathbf{x})+\mathbf{x}$, such that $(I_S \circ \mathrm{\Phi})(\mathbf{x})=I_T(\mathbf{x})$ for all $\mathbf{x} \in \Omega$.

The LVmyo registration in a cardiac cycle can be seen as a series of pairwise DIRs, each between a source image $I_S : \Omega \subset \mathbb{R}^3 \rightarrow \mathbb{R}$ and a target image $I_T : \Omega \subset \mathbb{R}^3 \rightarrow \mathbb{R}$. Each DIR aims to find a mapping $\mathrm{\Phi}: \mathbb{R}^3 \rightarrow \mathbb{R}^3$ in the form of $\mathrm{\Phi}(\mathbf{x})=u(\mathbf{x})+\mathbf{x}$, such that $(I_S \circ \mathrm{\Phi})(\mathbf{x})=I_T(\mathbf{x})$ for all $\mathbf{x} \in \Omega$.
The optimal registration between two images is found by optimization as

\begin{equation} \label{eq:opt_reg}
    \mathrm{\Phi} ^*=\underset{\mathrm{\Phi}}{\operatorname*{arg\,min}}\bigl\{ 
    \mathcal{L}_{\text{sim}}(I_S \circ \mathrm{\Phi}, I_T)
    +\lambda \mathcal{L}_{\text{reg}}(\mathrm{\Phi}) 
    \bigr\},
\end{equation}

where $\mathcal{L}_{\text{sim}}$ and $\mathcal{L}_{\text{reg}}$ are similarity and regularization losses, respectively. The similarity loss used for this project is the normalized cross correlation loss. As for the regularization loss we use the symmetric Jacobian determinant regularization defined by van Harten et al.~\cite{vanharten2024cycle} as

\begin{equation}
    \mathcal{L}_{\mathrm{sjac}}(\mathrm{\Phi}) = \min \left\{ \frac{\left(\operatorname{det} \nabla \mathrm{\Phi}-1\right)^2}{| \operatorname{det} \nabla \mathrm{\Phi} |}, \tau\right\},
\end{equation}

with $\nabla \mathrm{\Phi}$ as the Jacobian matrix of $\mathrm{\Phi}$ evaluated at point $\mathbf{x}$ and $\tau$ as a hyperparameter to clip the regularization penalty. As the van Harten et al.~\cite{vanharten2024cycle} suggest, we use $\tau=10$ in our work.

\subsection{Implicit Neural Representations}

The INR of a DIR is achieved using  a multi-layer perceptron (MLP) to parameterize the deformation field. The MLP takes a continuous coordinate $\mathbf{x}$ as input and outputs the corresponding deformation vector for the given point. Thus, the INR represents the deformation $\mathrm{\Phi}(\mathbf{x})=u(\mathbf{x})+\mathbf{x}$.

Based on the findings of Byra et al.~\cite{Byra2023INRnature} regarding the activation functions of INRs, we adopted the \texttt{SIREN}~\cite{sitzmann2019siren} architecture for modeling the MLP.
% Following the results of the investigative study of INRs' activation functions by Byra et al.~\cite{Byra2023INRnature} we model the MLP with the \texttt{SIREN}~\cite{sitzmann2019siren} architecture. 
The \texttt{SIREN} model utilizes sinusoidal activation functions in the network.  This defines the \textit{i}'th layer of the neural network as

\begin{equation}
    \phi_i(\mathbf{x}_i) = \sin \left( \omega \left( \mathbf{W}_i \mathbf{x}_i + \mathbf{b}_i\right) \right),
\end{equation}

where $\mathbf{W}_i$ and $\mathbf{b}_i$ are the trainable parameters of a linear layer, $\mathbf{x}_i$ is the input vector, and $\omega$ is a modulation hyper-parameter. We use $\omega=30$ as Byra et al.~\cite{Byra2023INRnature} suggest. A visualization of an INR with one hidden layer is shown in \cref{fig:data_network}.

\subsection{Signed Distance Fields}

The SDFs utilized in the registration process are derived from segmentation of the LVmyo using the TotalSegmentator~\cite{Wasserthal2023TotalSegmentator} framework. The SDFs provide more meaningful value over the whole image domain compared to the raw LVmyo segmentations.
They are generated using the physical spacing of the images to account for the differences in spacing over the different image dimensions. 
% The SDFs are thresholded at 20 mm from the LVmyo surface, and the positive values are scaled to the interval ${]}0,1{]}$, while the negative values are scaled to ${[}-1,0{[}$

For the registration process, the SDFs are defined similarly to the source and target images as $S_S : \Omega \subset \mathbb{R}^3 \rightarrow \mathbb{R}$ and $S_T : \Omega \subset \mathbb{R}^3 \rightarrow \mathbb{R}$. As the domain of the SDFs are the same domain as the images, the mapping between the source and target domains $\mathrm{\Phi}$ is also applied to the SDFs. 
% The SDFs are clipped at a distance of 20 mm from the LVmyo surface, thereafter they are scaled to the interval $[-1,1]$, while keeping the 0-level the same. The images containing HU-values are also linearly scaled to $[-1,1]$, such that the weighting between image and SDF is more interpretable. 

The optimal registration \cref{eq:opt_reg} is modified to include the SDFs

\begin{equation} \label{eq:opt_reg_sdf}
    \mathrm{\Phi} ^*=\underset{\mathrm{\Phi}}{\operatorname*{arg\,min}}\bigl\{
    (1-\alpha)\mathcal{L}_{\text{sim}}(I_S \circ \mathrm{\Phi}, I_T)
    + \alpha \mathcal{L}_{\text{sim}}(S_S \circ \mathrm{\Phi},S _T)
    +\lambda \mathcal{L}_{\text{reg}}(\mathrm{\Phi}) \bigr\}.
\end{equation}

The weighting between the SDFs and the images in the training process is controlled by the hyperparameter $\alpha$.

% The two approaches hold different promises, the movement in the sequential approach is smaller in each registration. However, the non-sequential approach regularizes the movement more, due to the 

\section{Experiments}

% \subsection{Registration Over a cardiac cycle}
% Two approaches for the registration of the LVmyo over a cardiac cycle are employed: sequential registrations and non-sequential registration. In the sequential approach, scan 0 is registered to scan 1, whereafter scan 1 is registered to scan 2, scan 2 to scan 3, and so forth. The non-sequential approach registers scan 0 to scan 1, scan 0 to scan 2, etc. 
% To make the two methods more comparable, the registrations go from scan 0 to scan 19, and the registrations do not loop back around.

Two approaches are used to register the LVmyo over a cardiac cycle: sequential and non-sequential registration. In the sequential approach, each frame is registered to the next ($f_0$ to $f_1$, $f_1$ to $f_2$, etc.). In the non-sequential approach, $f_0$ is directly registered to all subsequent frames ($f_0$ to $f_1$, $f_0$ to $f_2$, etc.). For comparability, both methods register frames from 0 to 19 without looping back.

Both approaches are primarily evaluated using the LVmyo segmentations of the target image and the transformed segmentations of the source image. This is done with the metrics; Dice Similarity Coefficient (DSC) and the $95\%$  Hausdorff Distance (HD95). For the single patient with annotated landmarks, we also evaluate the target registration error (TRE) in Euclidian distance.

\subsection{Experimental Details}
Each INR used for registration is parameterized with 5 hidden layers containing 256 neurons per layer.
The first network mapping from $f_0$ to $f_1$ is trained for 2000 epochs, subsequent networks are trained for 1000 epochs, each using the previously trained network as initialization.
The Training is conducted using the Adam optimizer~\cite{Kingma2014AdamAM} with a learning rate of $10^{-5}$ and a batch size of 10,000 points per epoch. The points are randomly sampled within a dilated mask of the whole heart, acquired from TotalSegmentator~\cite{Wasserthal2023TotalSegmentator}. For the loss \cref{eq:opt_reg_sdf} we use $\lambda =0.05$ for controlling the regularization and $\alpha=\left\{0.0, 0.8,1.0\right\}$ for experimentation with the weighting between SDF and CT scan in the loss. 
% These values have been experimentally derived.
The coordinate systems of both the SDFs and the CT frames are scaled to the interval of $[-1,1]^3$ following the procedure of Wolternik et al.~\cite{wolterink2021implicit}. 
All experiments were performed on an NVIDIA RTX A4000 GPU, where the training time for the first network in the sequence is $\sim 1$ minute  and $\sim 30$ seconds for subsequent networks. \footnote{Our code is publicly available at: \url{https://github.com/MMLowes/INR_reg_LVMyo}.}
% add code and training times

\subsection{Results}

\begin{figure}[t]
    \centering
    \includegraphics[width=\textwidth]{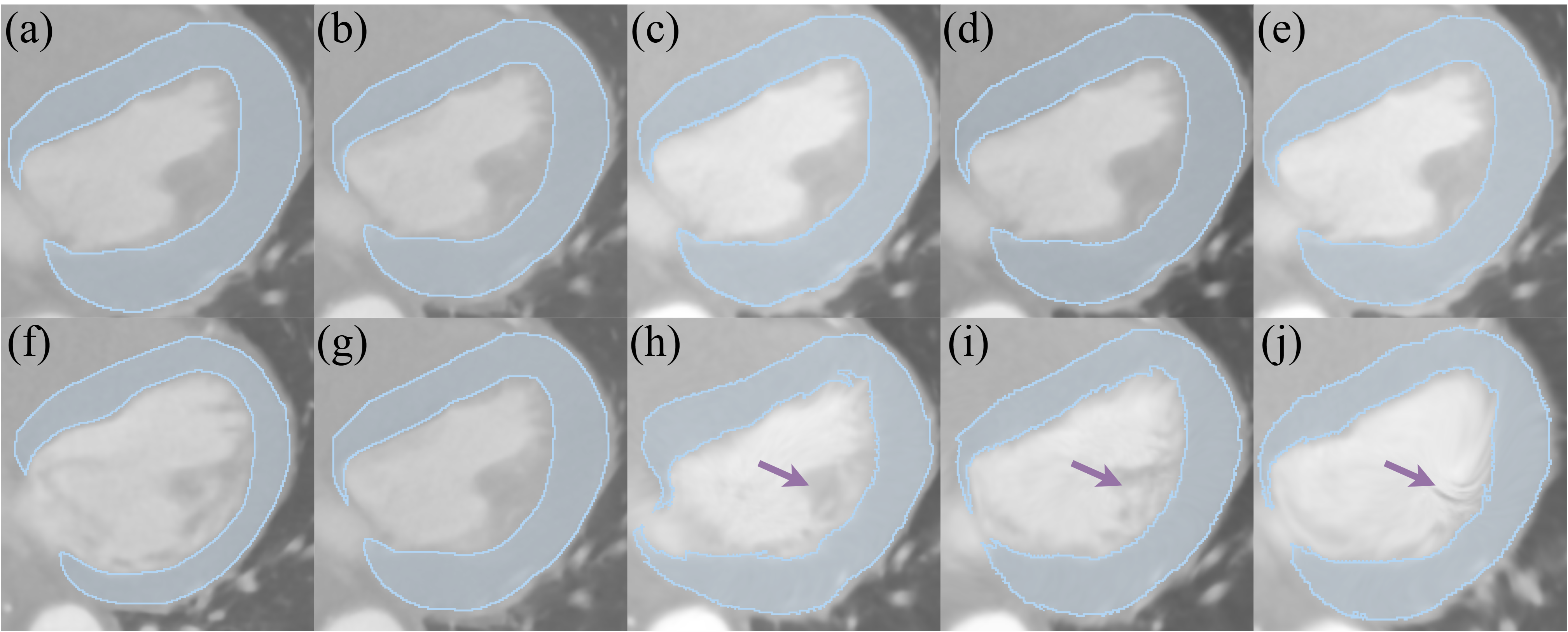}
    \caption{Axial slices of a registration example with the sequential approach in the top row (a) - (e) and the non-sequential approach in the bottom row (f) - (j).
    (a) and (b) shows the source and target frames which are before the ES-phase and the ES-phase it self, respectively. 
    (c) - (e) shows the transformed CT scan and LVmyo segmentation after the registration from (a) to (b) using $\alpha=0.0$, $\alpha=0.8$ and $\alpha=1.0$, respectively.
    (f) and (g)  shows the source and target frames which are before the ED-phase and the ES-phase.
    Again (h) - (j) shows the transformation  using using $\alpha=0.0$, $\alpha=0.8$ and $\alpha=1.0$, respectively.
    The arrows highlights trabeculated area of the LV.
    }
    \label{fig:reg_example}
\end{figure}

% \begin{figure}[t]
%     \centering
%     \includegraphics[width=\textwidth]{figs/reg_contrast.png}
%     \caption{Axial slices of a registration example with the non-sequential approach. (a) and (b) shows the CT frames at the ED- and ES-phase as well as the LVmyo segmentation. (c) - (e) shows the transformed CT frame and LVmyo segmentation after the registration from (a) to (b) using $\alpha=0.0$, $\alpha=0.8$ and $\alpha=1.0$, respectively.}
%     \label{fig:reg_example}
% \end{figure}

\setlength{\tabcolsep}{0.4em}
\begin{table}[t]
\caption{Evaluation of the two approaches with three different $\alpha$-values. All values are averaged over all time steps and the DSC and HD95 are also averaged over all patients.}
    \centering
    \begin{tabular}{lcccccc}
        \hline
        & \multicolumn{3}{c}{Sequential} & \multicolumn{3}{c}{Non-sequential} \\
        $\mathbf{\alpha}$-value & 0.0 & 0.8 & 1.0 & 0.0 & 0.8 & 1.0 \\ \hline
        \textbf{DSC} {[}\%{]}  & 94.79 & 97.78 & \textbf{98.31} &  91.18 & 96.90 & 97.28 \\ 
        \textbf{HD95} {[}mm{]} & 1.147 & 0.4364 & \textbf{0.3901} & 2.445 & 0.5212 & 0.4987 \\ 
        \textbf{TRE} {[}mm{]} & 11.40 & 8.703 & 11.03 & \textbf{4.739} & 6.273 & 8.347 \\ \hline
    \end{tabular}

    \label{tab:mean_results}
\end{table}

\begin{figure}[htb]
    \centering
    \includegraphics[width=\textwidth]{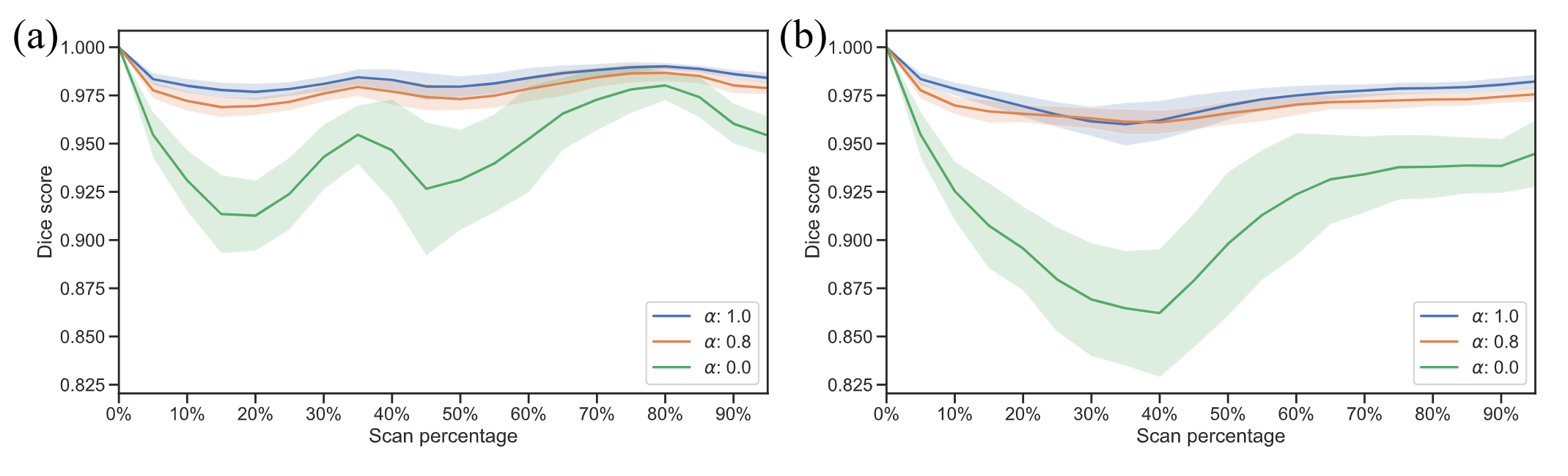}
    \caption{Evaluation between the LVmyo segmentation and the moved LVmyo segmentation averaged over all patients in DSC. (a)  shows the sequential approach, while (b) shows the non-sequential approach. The shaded area marks one standard deviation, and the scan percentage marks registration to this phase of the cardiac cycle.
    }
    \label{fig:dice_score}
\end{figure}

% A visualization of the same registration with different $\alpha$-values is shown in \cref{fig:reg_example}. 
% \Cref{fig:reg_example} illustrates a single registration across varying $\alpha$-values. The registrations are with the non-sequential approach from the ED-phase to the ES-phase, and highlight how the three different $\alpha$-values emphasizes the raw voxel values in the CT frame when $\alpha=0.0$, the LVmyo segmentation when $\alpha=1.0$ and a mixture when $\alpha=0.8$. 
\Cref{fig:reg_example} illustrates a single registration with both the sequential and non-sequential approaches across varying $\alpha$-values. The difference in the registration for the two approaches is from the source images (a) and (f), as the target images (b) and (g) are the same. The registrations between (a) and (b) in (c) - (e) are all very similar and of the same high qualtiy. However, the registration of (f) to (g) in (h) - (j) highlights how the three different $\alpha$-values emphasize the raw voxel values in the CT frame when $\alpha=0.0$, the LVmyo segmentation when $\alpha=1.0$ and a mixture when $\alpha=0.8$.

\Cref{tab:mean_results} shows the evaluation of the sequential and the non-sequential approaches to the registration task. Based on the DSC and HD95, it is indicated that training with $\alpha=1.0$ yields the best performance for both approaches.
However, the TRE of the method with $\alpha=1.0$ is worse in performance than when using $\alpha=0.8$.
The values for  DSC and HD95 in \cref{tab:mean_results} are averaged over both time and patients.
% These values are illustrated in \cref{fig:dice_score} as a function of the scan percentage.
The same values for DSC are illustrated  in \cref{fig:dice_score} as a function of the scan percentage, where the values are only averaged over patients.
This shows that in the non-sequential approach the registration accuracy is lower closer to the ES-phase of the cardiac cycle, mainly when using $\alpha=0.0$. The performance when using $\alpha=0.8$ and $\alpha=1.0$ are quite similar, there are no clear places where $\alpha=1.0$ outperforms $\alpha=0.8$ significantly.

% \begin{figure}[h]
%     \centering
%     \includegraphics[width=0.8\textwidth]{figs/dice_hausdorf.drawio.pdf}
%     \caption{Evaluation between the LVmyo segmentation and the moved LVmyo segmentation averaged over all patients in DSC (a), (b) and HD95 (c), (d). The shaded area marks one standard deviation. The scan percentage marks registration to this phase of the cardiac cycle.
%     (a) and (c) shows the sequential approach, while (b) and (d) shows the non-sequential approach.}
%     \label{fig:dice_score}
% \end{figure}

\begin{figure}[h]
    \centering
    \includegraphics[width=0.8\textwidth]{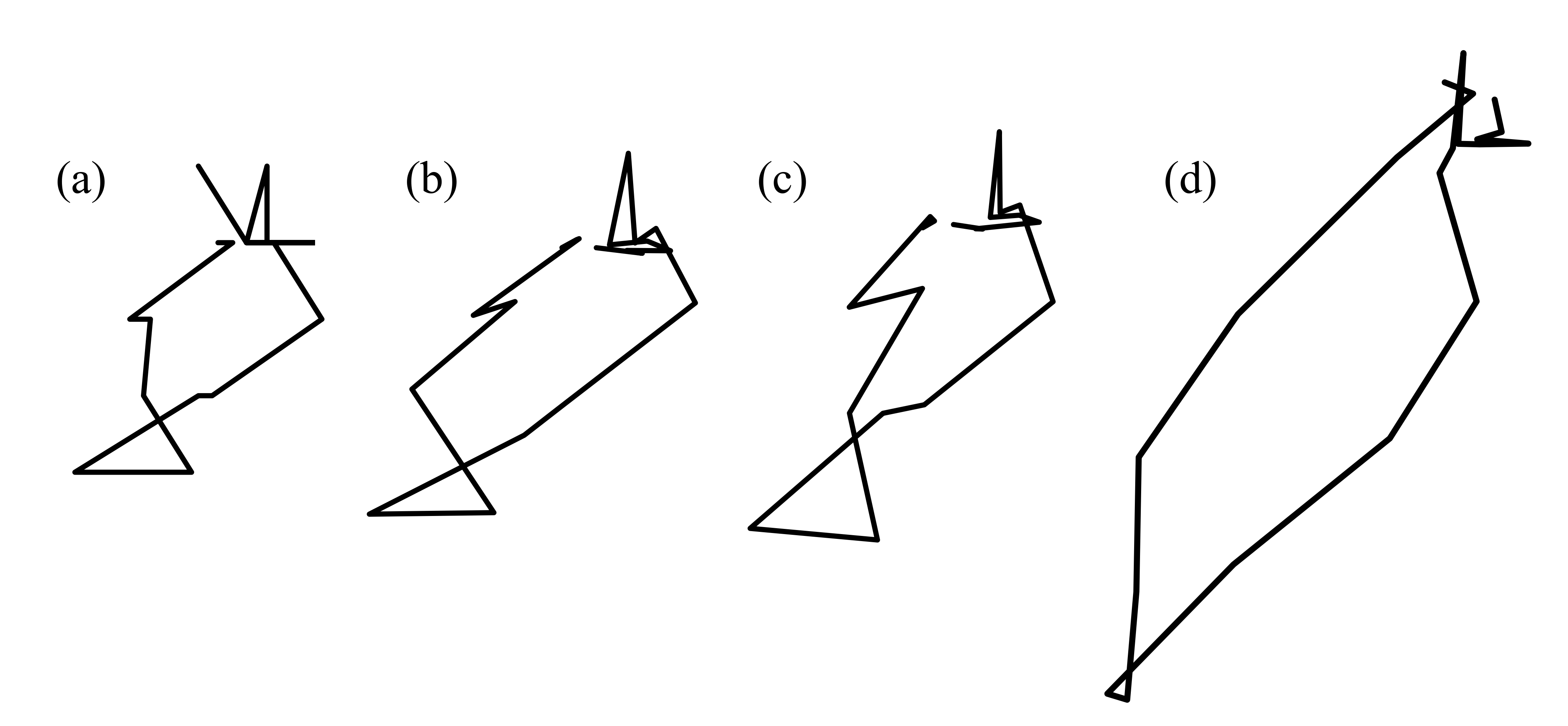}
    \caption{Movement of landmark 4 through a cardiac cycle illustrated in the sagittal plane. (a) is the annotated movement, while (b), (c) and (d) shows the movement using the non-sequential approach with $\alpha=0.0$, $\alpha=0.8$ and $\alpha=1.0$, respectively.}
    \label{fig:lm_track}
\end{figure}

From the patient with annotated landmarks, the movement of landmark 4 is visualized in \cref{fig:lm_track}. It is apparent that modeling the movement of the landmark with $\alpha=1.0$ differs from the annotated movement of the landmark. Meanwhile, modeling the movement with $\alpha=0.0$ and $\alpha=0.8$ follows the actual movement better.

% The distances of annotated landmarks to moved landmarks are illustrated in \cref{fig:landmarks}. The landmarks at the ED-phase are provided to the models, which then moves the landmarks through the cardiac cycle using the previously obtained registrations. 

% \begin{figure}[ht]
%     \centering
%     \includegraphics[width=0.9\textwidth]{figs/lanmarks_fig.drawio.pdf}
%     \caption{Distance between annotated landmarks and landmarks moved with the registrations. (a) shows the sequential approach, while (b) shows the non-sequential approach.}
%     \label{fig:landmarks}
% \end{figure}

\section{Discussion}
The sequential and the non-sequential approaches differ in how they handle registration. In the sequential approach, each registration step only needs to capture a small deformation. In contrast, the non-sequential approach handles increasingly larger deformations as the registration progresses closer to the ES-phase of the cardiac cycle, as the  movement is relative to the ED-phase. Consequently, we expect the sequential approach to achieve a higher DSC, as the registration should be easier for the INR to parameterize. However, the performance differences are not as substantial as anticipated, except for the models without the SDF. A difference of less than 0.5\% in DSC and 0.05 mm in HD95 between the sequential and the non-sequential approaches is a minor improvement. Indicating that the inclusion of SDF effectively addresses the challenge of larger deformations in the non-sequential approach.

The advantage of the non-sequential approach is seen in the landmark tracking. Modeling the movement of a set of points over a cardiac cycle with the sequential displacement of the points leads to a build-up of errors, which the non-sequential approach avoids as the displacement of the points are from the first frame each time. This build up of errors is what gives the high TRE in \cref{tab:mean_results}.
Although, this comparison is not strictly fair and only with landmarks from a single patient, it shows the major differences in the two approaches and why the sequential is not strictly better.

In general, the inclusion of SDFs in the registration makes a difference in the registration quality. This is apparent in \cref{fig:dice_score}, where both registrations with $\alpha > 0$ outperform the registration with $\alpha=0.0$, that is, without the SDFs. Including the original CT in the registration ($\alpha=0.8$) results in a slight drop in performance. However, this can potentially be attributed to the choice of evaluating only the LVmyo. The SDFs are based directly on the LVmyo segmentations, so aligning the SDFs perfectly gives the best score in both DSC and HD95. This alignment is possible without the CT frame, thus adding it will not improve performance with these metrics.
% However, evaluation on the whole heart segmentation from TotalSegmentator shows that $\alpha=0.8$ outperforms both other $\alpha$-values in the sequential and non-sequential approach at all times steps. 

It is known from echocardiography that the contraction of the LVmyo is a complex movement that also involves a twisting motion. This will not be faithfully captured if only using the segmentation masks or the derived SDF in the registration.
% The registration with only SDF is prone to an aperture problem. 
There are no direct constraints on the surface of the LVmyo, thus points can slide around on the surface without giving a larger similarity loss $\mathcal{L}_{\text{sim}}$. 
% This can explain why the landmark tracking is 
If this is happening, it is not shown in the DSC and HD95 metrics as the evaluation is only between segmentations.
When the CT frame is included in the registration the potential for modeling the twisting motion is greater, as there are more changes in the HU values on the surface. 
An example of this is what \cref{fig:lm_track} shows. The movement of this landmark is easier to accurately model when including HU values in the registration. This indicates that accurate modeling of the LVmyo is not feasible from only the SDF.
% The inclusion of the regularization term $\mathcal{L}_{\text{reg}}$ in the loss function constrains the amount that the points can slide on the surface at each time step. This potentially minimizes the problem, although accurate modeling of the LVmyo cannot be done from only the SDF.

% However, it is difficult to say how realistic the registrations with only SDFs are. The LVmyo makes a twisting motion in the contraction phase of the cardiac cycle. This twisting motion cannot be captured by the SDFs, thus SDFs are not able to model the exact movement of the LVmyo. 

In \cref{fig:reg_example}, examples of registrations are shown. Upon visual inspection of the non-sequential approach (f)-(j), the registration using solely SDF (j) achives the most accurate transformation of the LVmyo segmentation. However, the trabeculated area of the LV does not look as realistic as the other registrations (c) and (d), see arrows in figure. Using only the CT frame provides the best registration of the trabeculated area, but a worse registration of LVmyo segmentations. The combination of CT frame and SDF appears realistic in the trabeculated area of the LV, and the transformed LVmyo segmentation fits the target segmentation quite well. 

We have previously utilized the \texttt{elastix}~\cite{Klein2010elastix} framework for this type of registration. The performance of \texttt{elastix} lies around 60\% DSC on the LVmyo, when including the SDF in the registration from the ED-phase to the ES-phase. Another DIR method tried is the VoxelMorph~\cite{balakrishnan2019voxel} framework which achieves a performance of 80\% DSC in the same registration setup mapping from the ED-phase to the ES-phase when including the SDF as an extra feature of the input images. Both \texttt{elastix} and VoxelMorph have shorter inference times than the INR registration framework at $\sim 19$ seconds and $\sim 0.4$ seconds per image pair. However, the VoxelMorph framework needs training for $\sim 18$ hours on our setup, where the INRs are optimized during inference. Inspired by VoxelMorph we implemented the dice loss on TotalSegmentator segmentations to substitute the SDF. This yielded no significant improvement to using only HU values for the registration.
% TODO: integration of SDF as a feature channel.

% In the non-seqeuntial approach having both information from the CT scan and the SDF makes a substantial difference. The registrations with only SDFs seems to be stuck in a local minima in the ES phase of the cardiac cycle. Meanwhile the registrations based on only CT scans fail to capture more of the movement of the LVmyo. However, the combination of the two mediums achieves to capture the movement of the LVmyo.
% Consequently, this implies that the registrations with SDF are evaluated on what they are based on. 

\section{Conclusion}
We show that implicit neural representations are capable of modeling the movement of the left ventricle myocardium over a cardiac cycle. Additionally, the inclusion of the signed distance field in the registration process guides the registration to be more precise in the area of interest, specifically the left ventricle. Both the sequential and the non-sequential approaches show promising results and have potential for further development. Future work could include using both methods simultaneously to regularize each other as a form of cycle-consistency learning over a whole cardiac cycle.

\bibliographystyle{splncs04}
\bibliography{references.bib}

\begin{thebibliography}{10}
\providecommand{\url}[1]{\texttt{#1}}
\providecommand{\urlprefix}{URL }
\providecommand{\doi}[1]{https://doi.org/#1}

\bibitem{balakrishnan2019voxel}
Balakrishnan, G., Zhao, A., Sabuncu, M., Guttag, J., Dalca, A.V.: Voxelmorph: A learning framework for deformable medical image registration. IEEE TMI: Transactions on Medical Imaging  \textbf{38},  1788--1800 (2019)

\bibitem{Byra2023INRnature}
Byra, M., Poon, C., Rachmadi, M.F., Schlachter, M., Skibbe, H.: Exploring the performance of implicit neural representations for brain image registration. Scientific Reports  \textbf{13}(1) (Oct 2023). \doi{10.1038/s41598-023-44517-5}, \url{http://dx.doi.org/10.1038/s41598-023-44517-5}

\bibitem{Castillo2009DIRLAB}
Castillo, R., Castillo, E., Guerra, R., Johnson, V.E., McPhail, T., Garg, A.K., Guerrero, T.: A framework for evaluation of deformable image registration spatial accuracy using large landmark point sets. Physics in Medicine \& Biology  \textbf{54}(7),  1849--1870 (Apr 7 2009). \doi{10.1088/0031-9155/54/7/001}, epub 2009 Mar 5

\bibitem{harten2023deformable}
van Harten, L., Herten, R.L.M.V., Stoker, J., Isgum, I.: Deformable image registration with geometry-informed implicit neural representations. In: Medical Imaging with Deep Learning (2023), \url{https://openreview.net/forum?id=Pj9vtDIzSCE}

\bibitem{vanharten2024cycle}
van Harten, L.D., Stoker, J., I{\v{s}}gum, I.: Robust deformable image registration using cycle-consistent implicit representations. IEEE Transactions on Medical Imaging  \textbf{43}(2),  784--793 (2024). \doi{10.1109/TMI.2023.3321425}

\bibitem{Heinrich2022voxelPlusPlus}
Heinrich, M.P., Hansen, L.: Voxelmorph++ going beyond the cranial vault with keypoint supervision and multi-channel instance optimisation. In: Biomedical Image Registration: 10th International Workshop, WBIR 2022, Munich, Germany, July 10–12, 2022, Proceedings. pp. 85--95. Springer International Publishing, Cham (2022)

\bibitem{Kingma2014AdamAM}
Kingma, D.P., Ba, J.: Adam: A method for stochastic optimization. CoRR  \textbf{abs/1412.6980} (2014), \url{https://api.semanticscholar.org/CorpusID:6628106}

\bibitem{Klein2010elastix}
Klein, S., Staring, M., Murphy, K., Viergever, M.A., Pluim, J.P.: elastix: a toolbox for intensity-based medical image registration. IEEE Transactions on Medical Imaging  \textbf{29}(1),  196--205 (Jan 2010). \doi{10.1109/TMI.2009.2035616}, epub 2009 Nov 17

\bibitem{Rueckert1999FFD}
Rueckert, D., Sonoda, L.I., Hayes, C., Hill, D.L., Leach, M.O., Hawkes, D.J.: Nonrigid registration using free-form deformations: application to breast mr images. IEEE Transactions on Medical Imaging  \textbf{18}(8),  712--721 (Aug 1999). \doi{10.1109/42.796284}

\bibitem{sitzmann2019siren}
Sitzmann, V., Martel, J.N., Bergman, A.W., Lindell, D.B., Wetzstein, G.: Implicit neural representations with periodic activation functions. In: Proc. NeurIPS (2020)

\bibitem{Sokooti2017cnn}
Sokooti, H., de~Vos, B., Berendsen, F., Lelieveldt, B.P., I\v{s}gum, I., Staring, M.: Nonrigid image registration using multi-scale 3d convolutional neural networks. In: International Conference on Medical Image Computing and Computer-Assisted Intervention. pp. 232--239. Springer, Cham (September 2017)

\bibitem{Sun2022MIRNFMI}
Sun, S., Han, K., Kong, D., You, C., Xie, X.: Mirnf: Medical image registration via neural fields. ArXiv  \textbf{abs/2206.03111} (2022), \url{https://api.semanticscholar.org/CorpusID:249431729}

\bibitem{Tian2023NePhi}
Tian, L., Sengupta, S., Greer, H., Est{\'e}par, R.S.J., Niethammer, M.: Nephi: Neural deformation fields for approximately diffeomorphic medical image registration. ArXiv  \textbf{abs/2309.07322} (2023), \url{https://api.semanticscholar.org/CorpusID:261823163}

\bibitem{deVos2019cnnreg}
de~Vos, B.D., Berendsen, F.F., Viergever, M.A., Sokooti, H., Staring, M., Išgum, I.: A deep learning framework for unsupervised affine and deformable image registration. Medical Image Analysis  \textbf{52},  128–143 (Feb 2019). \doi{10.1016/j.media.2018.11.010}, \url{http://dx.doi.org/10.1016/j.media.2018.11.010}

\bibitem{Wasserthal2023TotalSegmentator}
Wasserthal, J., Breit, H.C., Meyer, M.T., Pradella, M., Hinck, D., Sauter, A.W., Heye, T., Boll, D.T., Cyriac, J., Yang, S., Bach, M., Segeroth, M.: {TotalSegmentator}: Robust segmentation of 104 anatomic structures in {CT} images. Radiology: Artificial Intelligence  \textbf{5}(5) (2023). \doi{10.1148/ryai.230024}, \url{https://doi.org/10.1148%2Fryai.230024}

\bibitem{wolterink2021implicit}
Wolterink, J.M., Zwienenberg, J.C., Brune, C.: Implicit neural representations for deformable image registration. In: Medical Imaging with Deep Learning 2022 (2022)

\bibitem{Wu2016Bspline}
Wu, Z., Lan, T., Wang, J., Ding, Y., Qin, Z.: Medical image registration using b-spline transform. International Journal of Simulation: Systems, Science \& Technology  \textbf{17},  1.1--1.6 (01 2016). \doi{10.5013/IJSSST.a.17.48.01}

\bibitem{Yang2017Patchbased}
Yang, X., Kwitt, R., Styner, M., Niethammer, M.: Quicksilver: Fast predictive image registration – a deep learning approach. NeuroImage  \textbf{158},  378–396 (Sep 2017). \doi{10.1016/j.neuroimage.2017.07.008}, \url{http://dx.doi.org/10.1016/j.neuroimage.2017.07.008}

\end{thebibliography}




% LNCS Homepage, \url{http://www.springer.com/lncs}, last accessed 2023/10/25
% \end{thebibliography}
\end{document}